\documentclass[aps,prl,showpacs,twocolumn,amsmath,amssymb,superscriptaddress,10pt]{revtex4-1}
\usepackage{graphicx}
\usepackage{dcolumn}
\usepackage{bm}
\usepackage{units}
\usepackage{upgreek}
\usepackage{ucs}
\usepackage{color}
\usepackage{hyperref}
\usepackage{braket}
\usepackage[all]{hypcap}
\preprint{APS/123-QED}

\begin{document}

\title{Non-equilibrium steady-states in a driven-dissipative superfluid}
\author{Ralf Labouvie}
\affiliation{Department of Physics and Research Center OPTIMAS, Technische Universit\"at Kaiserslautern, 67663 Kaiserslautern, Germany}
\affiliation{Graduate School Materials Science in Mainz, Staudinger Weg 9, 55128 Mainz, Germany}
\author{Bodhaditya Santra}
\affiliation{Department of Physics and Research Center OPTIMAS, Technische Universit\"at Kaiserslautern, 67663 Kaiserslautern, Germany}
\author{Simon Heun}
\affiliation{Department of Physics and Research Center OPTIMAS, Technische Universit\"at Kaiserslautern, 67663 Kaiserslautern, Germany}
\author{Herwig Ott}
\affiliation{Department of Physics and Research Center OPTIMAS, Technische Universit\"at Kaiserslautern, 67663 Kaiserslautern, Germany}
\email{ott@physik.uni-kl.de}

\begin{abstract}
We experimentally study a driven-dissipative Josephson junction array, realized with a weakly interacting Bose Einstein condensate residing in a one-dimensional optical lattice. Engineered losses on one site act as a local dissipative process, while tunneling from the neighboring sites constitutes the driving force. We characterize the emerging steady-states of this atomtronic device. With increasing dissipation strength $\gamma$ the system crosses from a superfluid state, characterized by a coherent Josephson current into the lossy site to a resistive state, characterized by an incoherent hopping transport. For intermediate values of $\gamma$, the system exhibits bistability, where a superfluid and a resistive branch coexist. We also study the relaxation dynamics towards the steady-state, where we find a critical slowing down, indicating the presence of a non-equilibrium phase transition.

\end{abstract}



\maketitle

Non-equilibrium steady-states constitute fix points of the phase space dynamics of classical and quantum systems \cite{Strogatz2014,Spohn1977,Gallavotti1995}. They emerge under the presence of a driving force and lie at the heart of transport phenomena such as heat conduction \cite{Schwab2000,Segal2003,Komatsu2008} or current flow \cite{Kohn1957,Pecchia2004,Novoselov2005}. They also naturally appear in open quantum systems \cite{Breuer2002,Daley2014} and are connected to the study of non-equilibrium thermodynamics and non-equilibrium quantum phase transitions \cite{DellaTorre2010}. It has been pointed out that engineering open quantum systems can induce a phase space dynamics which drives the quantum system in a pure state by solely dissipative means \cite{Diehl2008,Verstraete2009,Barreiro2011,Pedersen2014}. Controlling and understanding the non-equilibrium steady-states of an open many-body quantum system therefore offers new routes for quantum state engineering and out-of-equilibrium quantum dynamics. Here, we investigate the steady-states of a driven-dissipative Josephson junction array realized with a Bose-Einstein condensate in a one-dimensional optical lattice \cite{Cataliotti2001}. Varying the strength of the dissipation, the system can be tuned from superfluid to resistive transport. In between, it exhibits a region of bistability. The peculiar transport properties make such devices promising elements for complex atomtronic circuits. At the same time, they are an interesting candidate to study generic properties of an open quantum system. Our results manifest the high potential of open system control in ultracold quantum gases.
 
Open quantum systems are characterized by the competition between the intrinsic unitary dynamics, governed by the Hamilton operator $H$, and the coupling to the environment, which induces non-unitary time evolution and quantum jumps, described by jump operators $(\hat{a_i}$, $\hat{a_i}^\dagger)$ which act on the system with rates $\gamma_i$ \cite{Breuer2002}. The time evolution of the density matrix $\rho$ in Markov approximation is then described by a master equation in Lindblad form \cite{Lindblad1976}:

\begin{equation}
\dot{\rho}= \mathcal{L}(\rho) = -\frac{i}{\hbar}\left[H,\rho\right]+\sum_i\frac{\gamma_i}{2}\left( 2 \hat{a_i} \rho \hat{a_i}^\dagger - \hat{a_i}^\dagger\hat{a_i} \rho - \rho\hat{a_i}^\dagger\hat{a_i} \right).
\end{equation}

Non-equilibrium steady-states (NESS) are defined by the condition $\mathcal{L}(\rho_\mathrm{NESS})=0$. The steady-sate can be a mixed state or a pure state: $\rho_\mathrm{NESS}=\ket{\Psi_\mathrm{NESS}}\bra{\Psi_\mathrm{NESS}}$. When the jump operators do not affect a pure state, i.e. $\hat{a_i}\ket{\Psi_\mathrm{NESS}}=0$, the state $\ket{\Psi_\mathrm{NESS}}$ is called a dark state. Steady-states have the peculiar property that they can be attractor states of the phase space dynamics. Controlling an open quantum system can therefore be a robust way to prepare well-designed quantum states.

\begin{figure}[t]
\begin{center}
\includegraphics[scale=0.5]{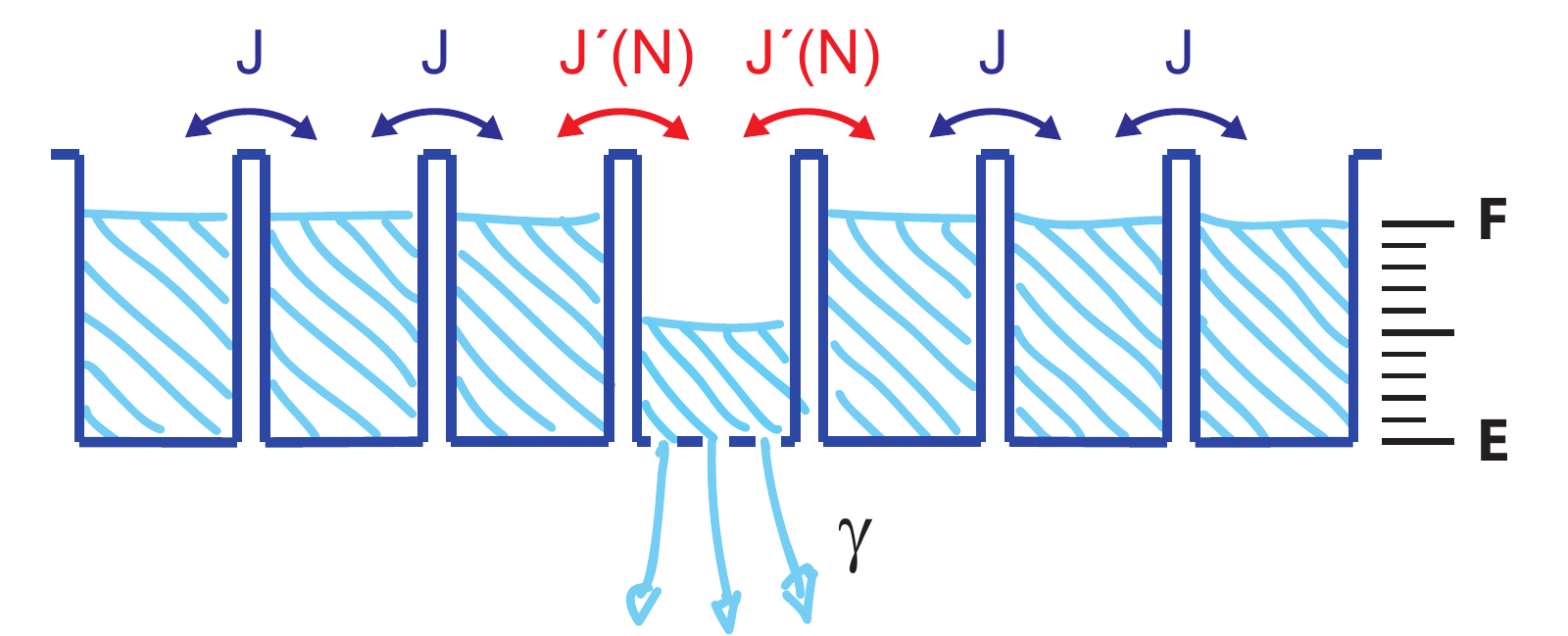}
\end{center}
\caption{Schematics of the experiment. One site of an array of superfluids is subject to an incoherent local loss process with rate $\gamma$. The coherent tunneling coupling between the reservoir sites is given by $J$. The coupling to the lossy site is given by $J'(N)$ and depends on the filling level (see text).}
\label{fig:1}
\end{figure}

In the experiment, we study a weakly-interacting, superfluid Bose-Einstein condensate of rubidium atoms in a one-dimensional periodic potential with high occupancy per site \cite{Cataliotti2001,Labouvie2014}. Each site contains a small condensate ($N_0\approx700$ atoms in the center of the trap) and all of them are connected via the tunneling coupling $J$, which is controlled by the height of the optical lattice. Employing a scanning electron microscopy technique \cite{Santra2015}, we introduce a well-defined local particle loss as a dissipative process in a single site of the system. To set the dissipation strength $\gamma$, we adjust the effective intensity of the electron beam. The corresponding jump operators are then given by the bosonic annihilation operators $\hat{a_i}$, acting on all spatial modes of the lossy site with the same dissipation rate $\gamma$. A fraction of the lost atoms is ionized by the electron beam and serves as a continuous probe of the occupation of the lossy site. The drive is provided by the large number of full sites left and right. The overall atom loss during a measurement is about 10\,\%, such that we can consider these sites as a superfluid reservoir. Fig.\,\ref{fig:1} shows a sketch of the experimental scenario. Related systems based on dissipative Bose-Hubbard models have been studied theoretically in Refs.\,\cite{Barmettler2011,Witthaut2011,Boite2013,Vidanovic2014}.

\begin{figure}[t]
\begin{center}
\includegraphics[scale=1.0]{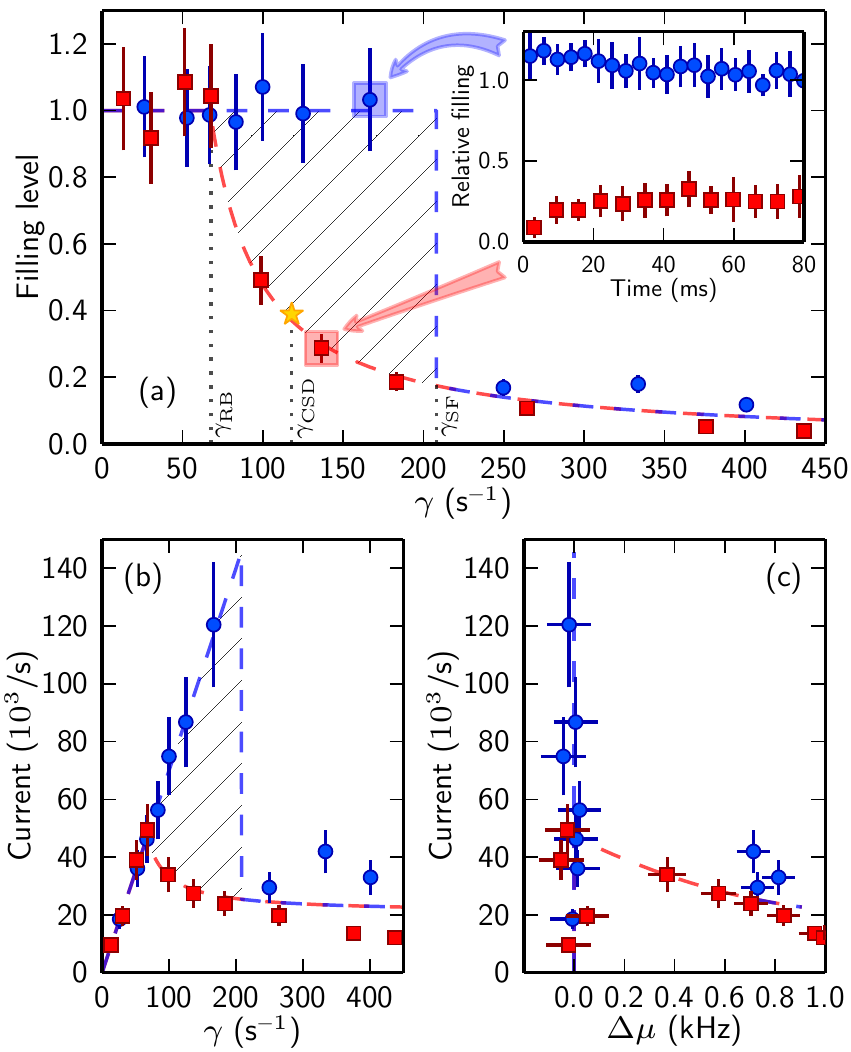}
\end{center}
\caption{(a) Steady-state filling level of the lossy site in dependence of the dissipation strength $\gamma$ for an initially full (blue circles) and empty (red squares) site. The tunneling coupling between the sites is $J/\hbar\;$=$\;\unit[230]{s^{-1}}$. The hatched area indicates the region of bistability (the dashed lines are a guide to the eye). $\gamma_{\mathrm{RB}}$, $\gamma_{\mathrm{CSD}}$ and $\gamma_{\mathrm{SF}}$ denote critical values of the dissipation and are explained in the main text. Inset: dynamical evolution of the system into the steady state within the bistable region for both initial conditions. (b) Steady-state current into the lossy site in dependence of $\gamma$. (c) For small values of $\Delta \mu$, the current-voltage characteristics display the typical behaviour of a superconductor.}
\label{fig:2}
\end{figure}

At the beginning of each experimental sequence, we initialize different starting conditions by optionally emptying the lossy site. Upon continuous dissipation, a steady-state is established on a time-scale of several tens of milliseconds. Once the system has reached the steady-state the losses are compensated by the refilling dynamics. At the end of the experimental sequence we freeze out and probe the final density distribution in a deep lattice. Fig. \ref{fig:2}(a) shows the resulting filling level of the lossy site in the steady-state in dependence of the dissipation rate $\gamma$. The two data sets correspond to an initially full site (blue points) and empty site (red points). For small values of $\gamma$, both initial conditions lead to a completely full site in the steady-state. For large dissipation, the lossy site is almost empty in both cases. The most prominent feature appears in between and is highlighted with the hatched area: the appearance of bistability. Starting from an empty site leads to a different filling level in the steady-state compared to starting from a full site. The inset in Fig. \ref{fig:2}(a) shows the two different trajectories, clearly displaying the two coexisting steady-states.

To further analyze the properties of the steady-states of the system we look at the current of atoms into the lossy site. The temporal evolution of the atom number in that site is given by
\begin{equation}
\label{Nt}
\dot{N}(t) = -\gamma N(t) + I(t)\, ,
\end{equation} 
where $N\left(t\right)$ is the number of atoms in the lossy site and $I\left(t\right)$ is the current from the reservoir sites. In the steady-sate ($\dot{N}=0$) we find $I_{\text{S}}=\gamma N_{\text{S}}$, where the subscript denotes the steady state value. This allows us to convert the filling level of Fig. \ref{fig:2}(a) into the current plot shown in Fig. \ref{fig:2}(b). For small dissipation, the current into the lossy site is exactly linear to the applied dissipation. This is quite remarkable, as the dissipation is externally applied and it is not at all obvious that the current response induced by the dissipation exactly balances the losses such that the site remains full. Because there is no difference in atom number between the sites in this regime, the current cannot be driven by a difference in chemical potential. Instead, it can only be driven by a quantum-mechanical phase gradient between the sites, thus constituting a supercurrent. What we observe is therefore the superfluid response of the system, which is capable to provide exactly the right current to counteract the losses. To visualize this effect we show in Fig. \ref{fig:2}(c) the current-voltage characteristic for the steady-states, converting atom number difference in chemical potential difference \cite{Labouvie2014}. The graph displays the well-known behaviour of a superconductor, where a supercurrent is present for vanishing voltage. 

The required build-up of a well-defined phase between adjacent sites can be derived from the discrete nonlinear Schr\"{o}dinger equation of coupled condensates \cite{Cataliotti2001}:
\begin{align}
i\hbar \frac{\partial \psi_n}{\partial t} = -J & \left(\psi_{n-1}+\psi_{n+1}\right)+U\left|\psi_n\right|^2\psi_n\\ \notag
&+i\frac{\gamma_{\text{dis}}}{2}\psi_n\delta_{nm}\; ,
\end{align}
where $J$ is the tunneling coupling, $U$ the on-site interaction and $m$ denotes the lossy site. In this meanfield version of the problem, the losses are implemented as an imaginary potential \cite{Barontini2013}. The model has a steady-state solution with unity filling at each site and a phase difference of $\sin (\Delta\Phi)=\hbar\gamma/(4J)$ between all adjacent sites. This exemplifies how a completely incoherent process can induce a well-defined quantum mechanical phase. The theoretical model also suggests that the steady-state is a pure state, corresponding to a Bloch state with finite quasi momentum $q$. This Bloch state is an attractor state of the phase space dynamics and its generation is a fundamental example for dissipative quantum state engineering. A similar situation has been theoretically studied for a three well system in Ref.\,\cite{Burnett1999}. There it was found that dissipation and interactions lead to a well defined relative phase between the three wells.

The superfluid response of an initially full site breaks down above a critical dissipation strength $\gamma_{\mathrm{SF}}\approx J/\hbar $. The above meanfield treatment predicts steady-states with a four times higher maximum superfluid current for a dissipation strength of $4J/\hbar$. We attribute this difference to the decoherence which is induced by the particle loss but is not accounted for in the meanfield treatment. The observation of a supercurrent for $\gamma < \gamma_{\mathrm{SF}}$ is analog to a supercurrent in a voltage biased Josephson junction \cite{Steinbach2001}. We therefore refer to this class of steady-states as the superfluid branch. 

\begin{figure}[b]
\begin{center}
\includegraphics[scale=1.0]{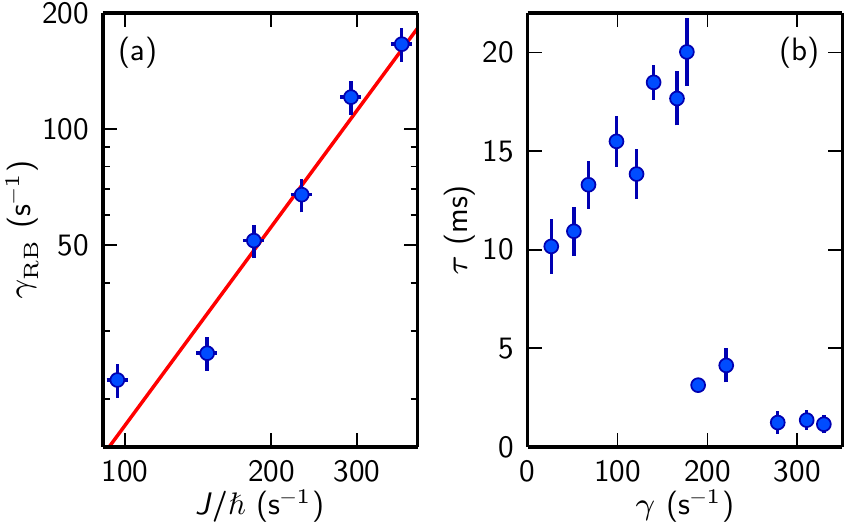}
\end{center}
\caption{(a) Critical dissipation rate $\gamma_\mathrm{RB}$ in dependence of the tunneling rate. We find a power law dependence with an exponent 1.7(2). This corresponds to a transition from a coherent to a incoherent process at which internal rates become proportional to $J^2$. (b) Time $\tau$ in which the steady state is reached for different dissipation rates at $J/\hbar\;$=$\;\unit[290]{s^{-1}}$. Within the bistable region $\tau$ is increasing which is known as critical slowing down.}
\label{fig:3}
\end{figure}

Starting from an empty site leads to a more complex behaviour as the multimode structure provided by the radial degrees of freedom in each site becomes important. The refilling dynamics in the absence of dissipation scales quadratically with $J$, which is characteristic for an incoherent transport process \cite{Labouvie2014}. In the presence of dissipation we find that below a critical dissipation rate $\gamma_\mathrm{RB}$, the lattice site refills completely, reaching again a superfluid steady-state as discussed above (Fig. \ref{fig:2}(a)). This highlights the fix point character of the steady-state, which is established independent from the initial conditions. The scaling of $\gamma_\mathrm{RB}$ with $J$ follows a power law with an exponent of 1.7(2) (Fig. \ref{fig:3}(a)). This is compatible with the quadratic dependence observed for the bare refilling. The dissipation thus initially competes with a classical hopping process. When the dissipation rate exceeds $\gamma_\mathrm{RB}$, the site does not fill up completely and the steady-state is characterized by a difference in atom number. This corresponds to a finite difference in chemical potential, in analogy to a voltage drop in a conductor. We therefore refer to this branch as the resistive branch (RB), where the transport can no longer be provided by a supercurrent alone, but must at least include an incoherent part or is even fully incoherent. 

Between the two critical dissipation strengths $\gamma_{\mathrm{SF}}$ and $\gamma_\mathrm{RB}$, the system exhibits bistability. Bistability occurs in various physical systems, e.g. in nonlinear optics \cite{Drummond1980} and electronic tunneling devices \cite{Goldman1987}. Closely related to our experiment is the appearance of a superfluid branch and a dissipative branch in biased Josephson junctions \cite{Schoen1990}. In this work, the microscopic origin of the bistability is a nonlinear tunneling coupling $J'(N)$, which depends on the difference in atom number between the lossy site and the superfluid reservoir. Due to the interaction energy in each site, tunneling is limited to energetically matching radial states of the empty site leading to an additional Franck-Condon factor \cite{Labouvie2014}. Thus, an increasing population difference results in a suppression of tunneling, eventually causing bistability in the system. The observation of bistability in a many-body quantum system might come as a surprise as a quantum system can in principle tunnel from one steady-state to another one. In the present situation however, the high occupancy makes such a tunnel process unlikely. In future investigations of Bose-Hubbard systems, tunneling between bistable steady-states might become observable.   

An important question that occurs in the dynamics of this Josephson device is the nature of the phase transition from the resistive to the superfluid behaviour. Starting from an initially empty site (red points in Fig.\,2a) we have measured the timescale $\tau$ in which the steady-state is established. In Fig. \ref{fig:3}(b), we show $\tau$ in dependence of the dissipation rate. Increasing the dissipation rate, $\tau$ increases accordingly, reaching a maximum value and drastically dropping at a critical dissipation strength $\gamma_\mathrm{CSD}$. This is reminiscent of a critical slowing down \cite{Bonifacio1979} and appears well within the bistable region (indicated with a star in Fig. \ref{fig:2}(a)). We assume that this critical slowing down originates form the phase transition from a normal gas to an out-of-equilibrium condensate in the lossy site. Without dissipation, the conditions for Bose-Einstein condensation are already reached for a filling level of 10\,\% and one would expect the appearance of a condensate throughout the resistive branch. In the presence of dissipation however, the formation of a condensate competes with the losses, thus modifying or even inhibiting a phase transition. Two experimental observations support the presence of a phase transition. (i) The current in the resistive branch for $\gamma<\gamma_\mathrm{CSD}$ is higher than any current we observe for the bare refilling dynamics without dissipation. This suggests a partially superfluid transport. (ii) The width of the radial density distribution in the lossy site is slightly reduced for $\gamma<\gamma_\mathrm{CSD}$ despite the higher atom number. This suggests a condensation process in the lossy site. Because all involved timescales, the dissipation rate $\gamma$, the effective tunneling coupling $J'(N)$ and the collision rate \cite{Labouvie2014} are of similar magnitude, such a condensation process is likely to be a non-equilibrium process, as observed, e.g., in exciton-polariton condensates \cite{Richard2005}. In the future, a detailed study of the atom number fluctuations and the scaling laws around $\gamma_\mathrm{CSD}$ will offer new ways to characterize and classify such non-equilibrium phase transitions with a high level of control. 

\begin{figure}[t]
\begin{center}
\includegraphics[scale=1.0]{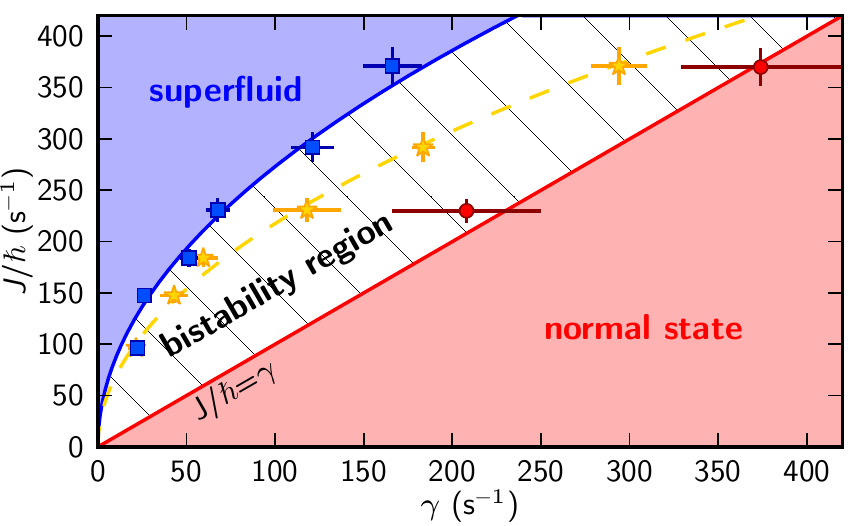}
\end{center}
\caption{Phase diagram of the steady-states. The blue shaded area marks the region where the lossy site is always filled independent from the initial condition and a supercurrent is induced. The red shaded area denotes the region of normal steady-states, where the system ends up in an almost empty site with an incoherent hopping transport, driven by the difference in chemical potential. In between (hatched area) the system is bistable, resulting in different steady-states depending on the initial conditions. The blue squares (red dots, yellow stars) denote $\gamma_\mathrm{RB}$ ($\gamma_\mathrm{SF}$, $\gamma_\mathrm{CSD}$). The blue line is the fit from Fig. \ref{fig:3}(a), the yellow line is a guide to the eye and the red line marks the boundary, where the dissipation strength $\gamma$ equals the coherent coupling $J/\hbar$.}
\label{fig:4}
\end{figure}

All previously obtained results can be summarized in a phase diagram (Fig. \ref{fig:4}), where the steady-states are classified in three regimes. For small dissipation rates and high tunneling couplings the system ends up in a superfluid state (blue shaded area) independent from the starting condition. Adjacent is the bistable region (hatched area) in which the system stays superfluid starting from a full site but ends up in a steady-state with finite atom number difference when starting from an empty site. Further increasing the dissipation rate connects the bistable region to the normal region (red shaded area) characterized by low filling and incoherent hopping transport. The data points are the corresponding boundaries for $\gamma_\mathrm{RB}$, $\gamma_\mathrm{CSD}$, and $\gamma_\mathrm{SF}$  obtained from the analysis of Fig. \ref{fig:2}(a) and Fig. \ref{fig:3}(b) for different tunneling couplings respectively. Within the bistable region, we find a critical slowing down (indicated by the yellow line), which we interpret as an out of equilibrium condensation process in the resistive branch. In Ref.\,\cite{DellaTorre2010}, it has been studied how non-equilibrium noise affects the normal to superconductor phase transition in a single Josephson junction, which is described by a simple phase boundary. In the present work, the nonlinear tunneling coupling prevents the existence of a simple phase boundary, but introduces a region of bistability between the superfluid and normal steady-states. The rich phase diagram promotes the dissipative Josephson junction as a versatile tool for atomtronics applications. The transport properties can be switched betwene different regimes and tuned over a large parameter range. They can even be varied dynamically, thus taking advantage of the intrinsic hysteresis in the system.

Engineering a local loss process in an ultracold quantum gas, we have achieved a high level of control over an open many-body quantum system. This has allowed us to study and characterize the steady-state phase diagram of a driven dissipative superfluid. In the future, studying the fluctuations around the steady states will be a tool to look at generalized dissipation fluctuations theorems for non-equilibrium systems \cite{Prost2009}. The exploration of quantum phases with help of competing dissipation mechanisms \cite{Lang2014}, the quest for complex many-body dark states \cite{Diehl2008,Krauss2008,Verstraete2009,Diehl2011,Kordas2012} and the investigation of non-equilibrium phase transitions \cite{Brennecke2013} make open system control a paradigm for future quantum research.

\begin{acknowledgments}
We thank Michael Fleischhauer, Eugene Demler, Axel Pelster, Sandro Wimberger and Jacob Sherson for fruitful discussions. This work was supported by the Deutsche Forschungsgemeinschaft within the Graduate School of Excellence MAINZ.
\end{acknowledgments}

\bibliographystyle{apsrev4-1}
\bibliography{dissipation}

\end{document}